\documentclass[a4paper,11pt]{article}
\usepackage{pos}
\usepackage[normalem]{ulem}
\usepackage{enumitem}

\title{Deciphering the role of multiple scatterings and time delays in the in-medium emission process}
 \ShortTitle{Multiple scatterings and time delays in medium-induced radiation}

\author[a]{Carlota Andres}
\author[b,c]{Liliana Apolin\'ario}
\author[d]{Fabio Dominguez}
\author*[d]{Marcos Gonzalez Martinez}
\author[d]{Carlos A. Salgado}

\affiliation[a]{CPHT, CNRS, Ecole Polytechnique,\\
IP Paris, F-91128 Palaiseau, France}

\affiliation[b]{LIP,\\
Av. Prof. Gama Pinto, 2, P-1649-003 Lisboa, Portugal}

\affiliation[c]
{Instituto Superior Técnico (IST), Universidade de Lisboa,\\
Avenida Rovisco Pais 1, 1049-001 Lisbon, Portugal}

\affiliation[d]{Instituto Galego de F\'isica de Altas Enerx\'ias IGFAE, Universidade de Santiago de Compostela,\\
Santiago de Compostela (Galicia), Spain}

\emailAdd{carlota.andres-casas@polytechnique.edu}
\emailAdd{liliana@lip.pt}
\emailAdd{fabio.dominguez@usc.es}
\emailAdd{marcosg.martinez@usc.es}
\emailAdd{carlos.salgado@usc.es}

\abstract{In this work we use the all-order resummed solution of the BDMPS-Z spectrum to shed light on the dynamics that controls the in-medium radiation process for each kinematical regime. We find that multiple scatterings are essential to correctly describe the radiation process both in the low and mid-energy regime, while in the high-energy region one single hard scattering is enough. Furthermore, we compute the all-order spectrum when the medium is produced with a time delay with respect to the hard process in which the parent parton was created. The propagation of the hard parton though vacuum before the medium formation induces extra medium-induced radiation which might have an impact on phenomenological analyses.}

\FullConference{%
  *** Particles and Nuclei International Conference - PANIC2021 ***\\
  *** 5 - 10 September, 2021 ***\\
  *** Online ***
}

\begin{document}
\maketitle

\section{Introduction}
The in-medium single inclusive gluon emission spectrum is the basic ingredient of medium-induced radiation calculations.  Given that its full numerical evaluation including all multiple scatterings has been cumbersome, analytical approximations, such as the harmonic oscillator (HO) and the opacity expansion, are usually employed in order to evaluate this spectrum. In this work we employ a recent framework that allows us to numerically compute the all-order BDMPS-Z spectrum without any additional assumptions \cite{Andres:2020vxs} to determine which is the range of validity of the analytical approaches and which physical mechanisms dominate the radiation process across all  energy regimes. In addition, we re-derive the all-order single gluon emission spectrum for a medium produced with a time delay with respect to the hard process in order to gain insight into the role played by the initial stages in the radiation process.

\section{Energy regimes of the medium-induced radiation spectrum}

In this section, we analyze the single gluon radiation spectrum across all energy regions. We look first at the low-energy region, also known as Bethe-Heitler regime, where the fully resummed spectrum off a highly energetic parton traversing a medium of length $L$ and linear density of scattering centers $n_0$ is given by \cite{Andres:2020kfg}
\begin{equation}
\left.\omega\frac{dI^{\mathrm{med}}}{d\omega}\right|_{\omega\to 0} = \frac{2\alpha_s C_R}{\omega} \Re\int_0^L ds\; n_0 \int_0^s dt\int_{\vec{p}\vec{q}}i\,\frac{\vec{p}\cdot\vec{q}}{\vec{q}^2}\sigma(\vec{q}-\vec{p})e^{-\left(i \frac{p^2}{2\omega}+\frac{1}{2}n_0\Sigma(p^2)\right)(s-t)}\,,
\label{eq:speclowomega}
\end{equation}
where $\omega$ is the  energy of the radiated gluon and $\Sigma$ is the factor introduced in Eq.~(4.4) of \cite{Andres:2020kfg}. The details of the interaction between the parton and the medium are encoded in the so-called dipole cross section $\sigma$ and throughout this work a Yukawa interaction model will be used.\footnote{We refer the reader to Ref.~\cite{Andres:2020kfg} for further details and for the derivation of Eq.~(\ref{eq:speclowomega}).} 

\begin{figure}[h]
\centering
\includegraphics[width=0.41\textwidth]{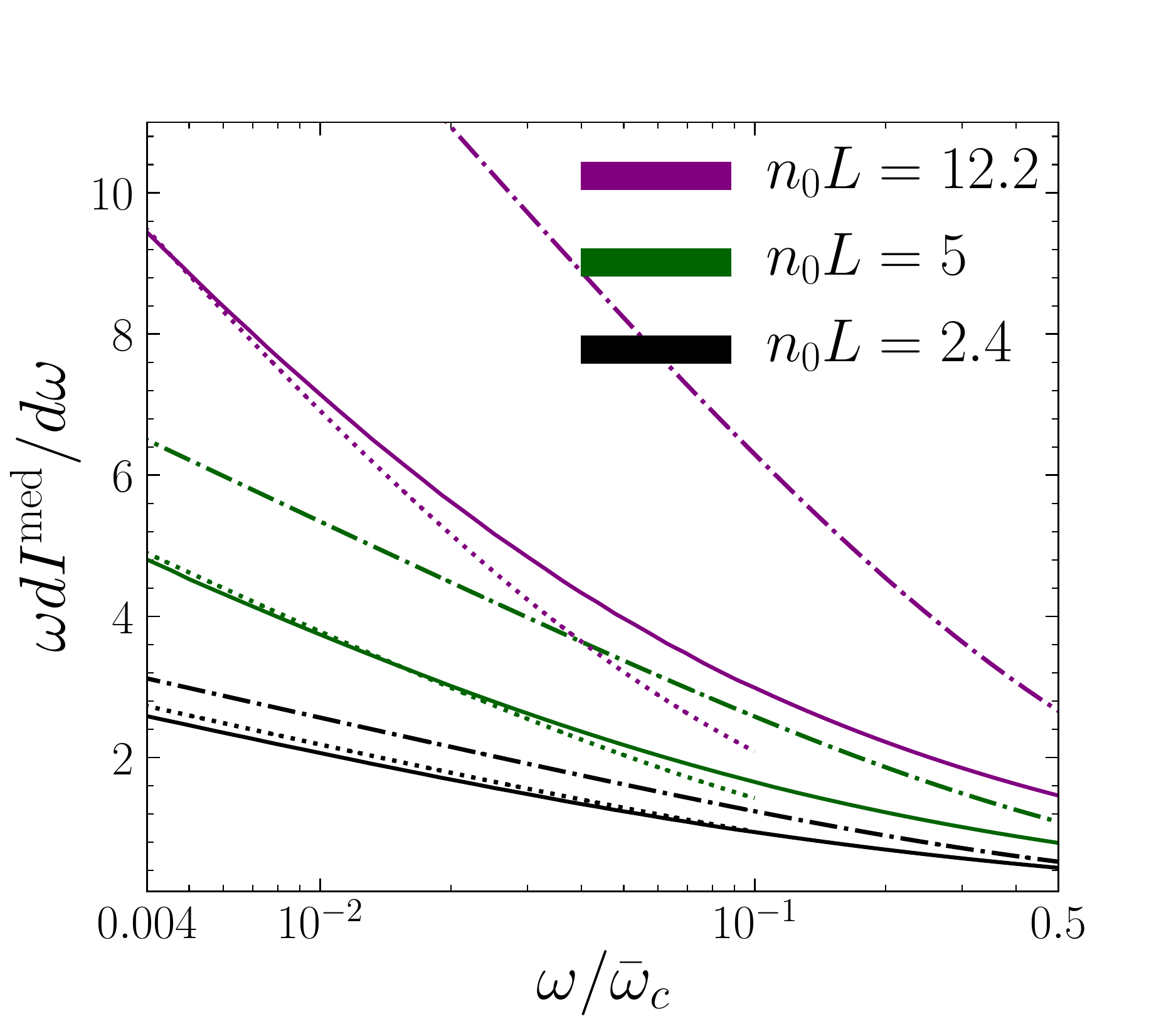}
\caption{All-order medium-induced gluon energy spectrum for a Yukawa-type interaction (solid lines), its low-energy asymptotic limit given by Eq.~(\ref{eq:speclowomega}) (dotted), and the GLV $N=1$ approximation (dash-dotted) as function of $\omega/\bar{\omega}_c=2\omega/\mu^2L$ ($\mu$ being the screening mass of the interaction) for different values of $n_0L$.} 
\label{fig:low_energy}
\end{figure}

One can easily notice that Eq.~(\ref{eq:speclowomega}) has the form of the first opacity result times a suppression factor accounting for the probability of not experiencing any further scatterings. Since this factor arises from the resummation of the virtual contributions \cite{Andres:2020kfg}, it shows that the inclusion of multiple scatterings is imperative in order to correctly compute the single gluon emission spectrum in this regime. This is shown in Fig.~\ref{fig:low_energy}, where we compare the low-energy result given by Eq.~(\ref{eq:speclowomega}) with the fully resummed and first opacity ($N=1$ GLV) results for a Yukawa-type interaction in the low energy regime. We observe that the single scattering approximation ($N=1$ GLV) overpredicts the all-order energy spectrum, whereas the asymptotic expression given by Eq.~(\ref{eq:speclowomega}) is a good approximation of the full result for sufficiently low energies, thus confirming that the inclusion of the above-mentioned suppression factor, and hence of multiple scatterings, is essential to properly describe the emission process in this region.

\begin{figure}[h]
\centering
\includegraphics[width=0.8\textwidth]{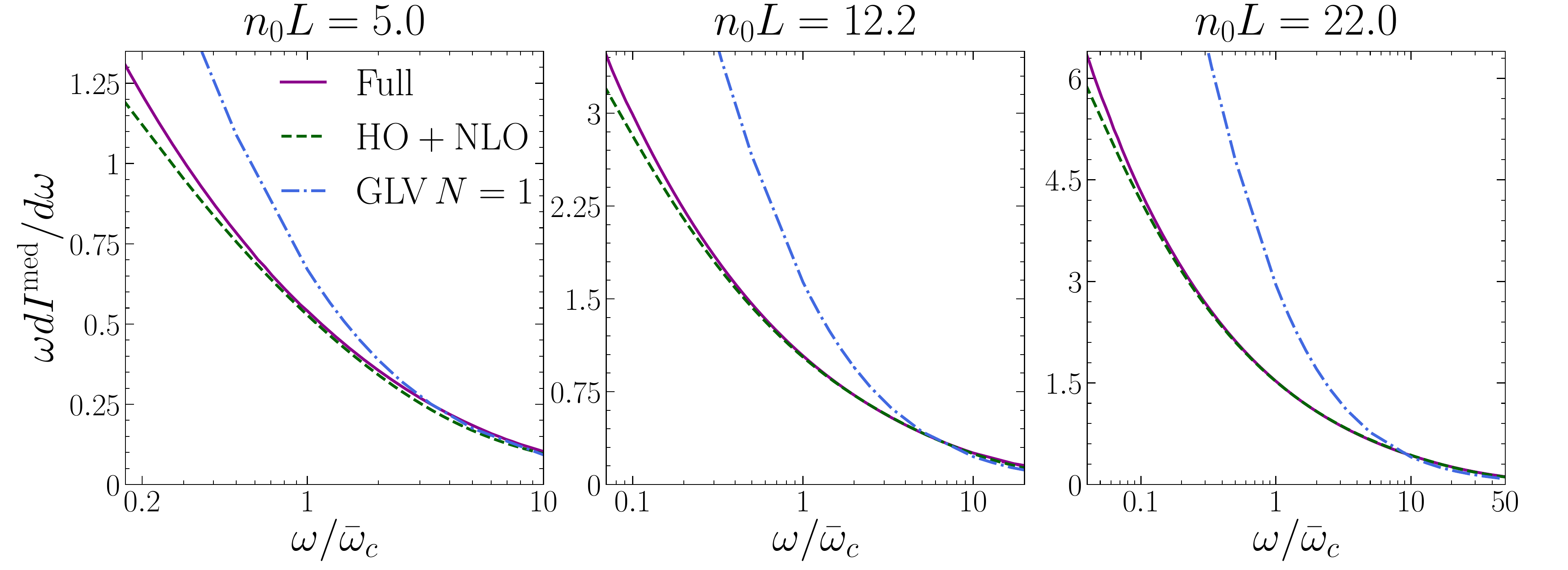}
\caption{All-order in-medium gluon energy spectrum for a Yukawa-type interaction (purple solid lines) \cite{Andres:2020vxs} compared to the first opacity (GLV $N=1$) (blue dash-dotted) and the IOE HO+NLO (green dashed) \cite{Mehtar-Tani:2019tvy} results as a function of  $\omega/\bar{\omega}_c=2\omega/\mu^2L$ for several values of $n_0L$.}
\label{fig:spectra}
\end{figure}

We move now to the intermediate- and high-energy regimes, where we compare the fully resummed energy spectrum with the first opacity and improved opacity expansion (IOE) results \cite{Mehtar-Tani:2019tvy}.  While the former is based on the assumption that the emission process is dominated by one single hard scattering, the latter consists of an expansion around the HO approximation, hence accounting for multiple scatterings. As we can see in Fig.~\ref{fig:spectra}, at intermediate gluon energies the first opacity approximation (dashed-dotted lines) overpredicts the fully resummed spectrum (solid), while the IOE (HO+NLO) (dashed) is a good approximation of the all-order solution. This manifests that the radiation process in this region cannot be correctly described without including multiple scatterings. When moving to higher energies ($\omega > \bar \omega_c$) both the fully resummed and IOE calculations converge to the GLV spectrum showing that, as expected, the emission process is governed by one single hard scattering in this region.



\section{Including a time delay in the production of the medium}

\begin{figure}[h]
\centering
\includegraphics[width=0.23\textwidth]{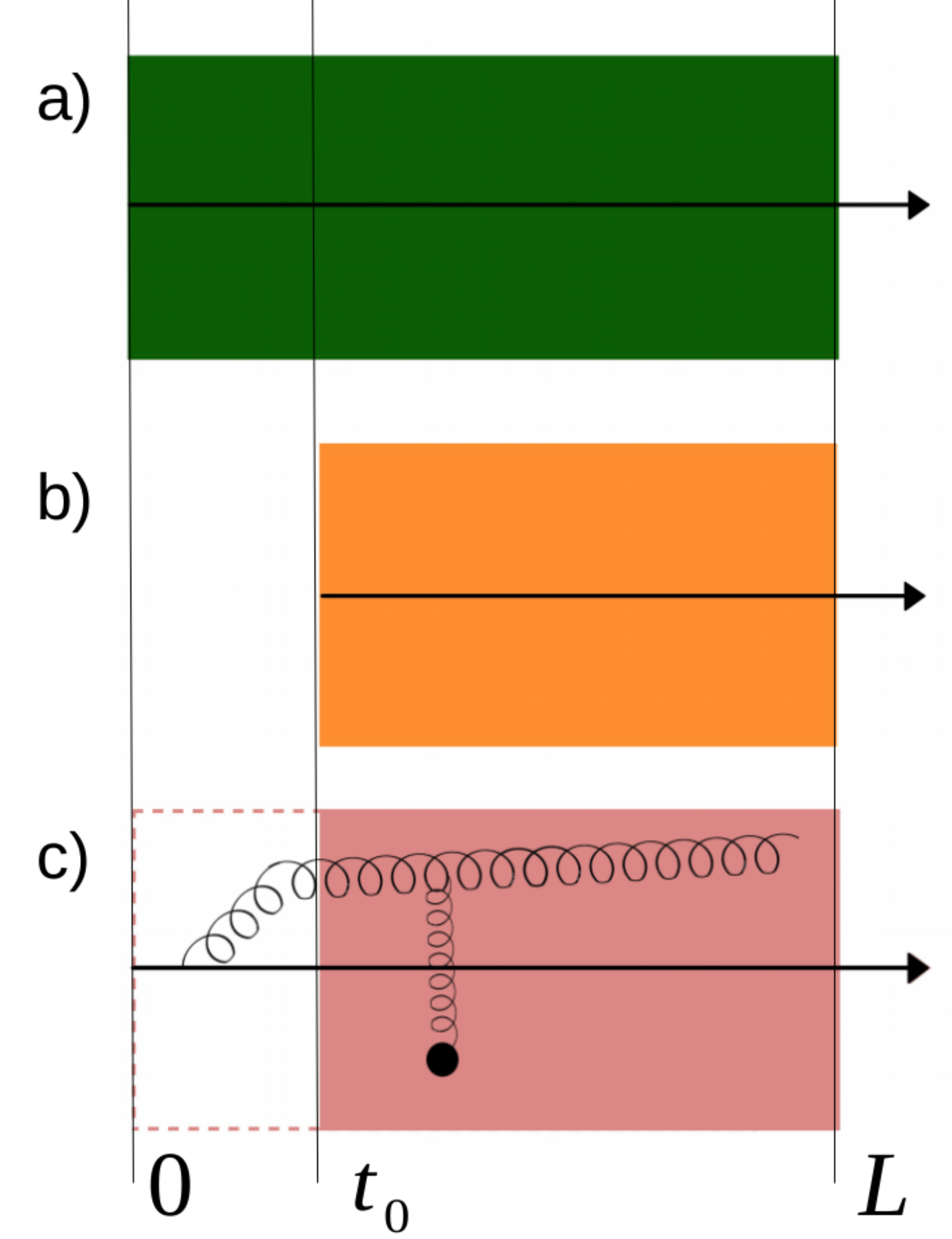}
\caption{Schematic representation of the three different scenarios analyzed.} 
\label{fig:cases}
\end{figure}
The theoretical calculations of the single gluon in-medium spectrum (as the ones presented in the previous section) are performed assuming a thermalized medium of length $L$. More precisely, the parent parton is set to be produced at a time $0$ and propagates through a medium going from $0$ to $L$ -- see case a) of Fig.~\ref{fig:cases}. In order to perform phenomenological studies, this kind of calculations are  embedded into more realistic environments where the medium properties are extracted from hydrodynamic simulations. For times prior to the applicability of hydrodynamics, usually no jet quenching is assumed. In practice, this corresponds to setting the parent parton and the medium to be produced at a time $t_0$, corresponding to the hydrodynamization time of the hydrodynamic model, as shown is case b) of Fig.~\ref{fig:cases}). However, the parent parton should be produced in the initial hard process, and thus witness the full evolution of the system including the initial stages prior to hydrodynamization.
In order to take into account the propagation of the hard parton in the initial stages (prior to $t_0$), we will assume the hard parton to be created at time $0$ propagating trough vacuum until $t_0$, and then through a thermalized medium from $t_0$ to $L$ as shown in case c) of Fig.~\ref{fig:cases}. 

\begin{figure}[h]
\centering
\includegraphics[width=0.41\textwidth]{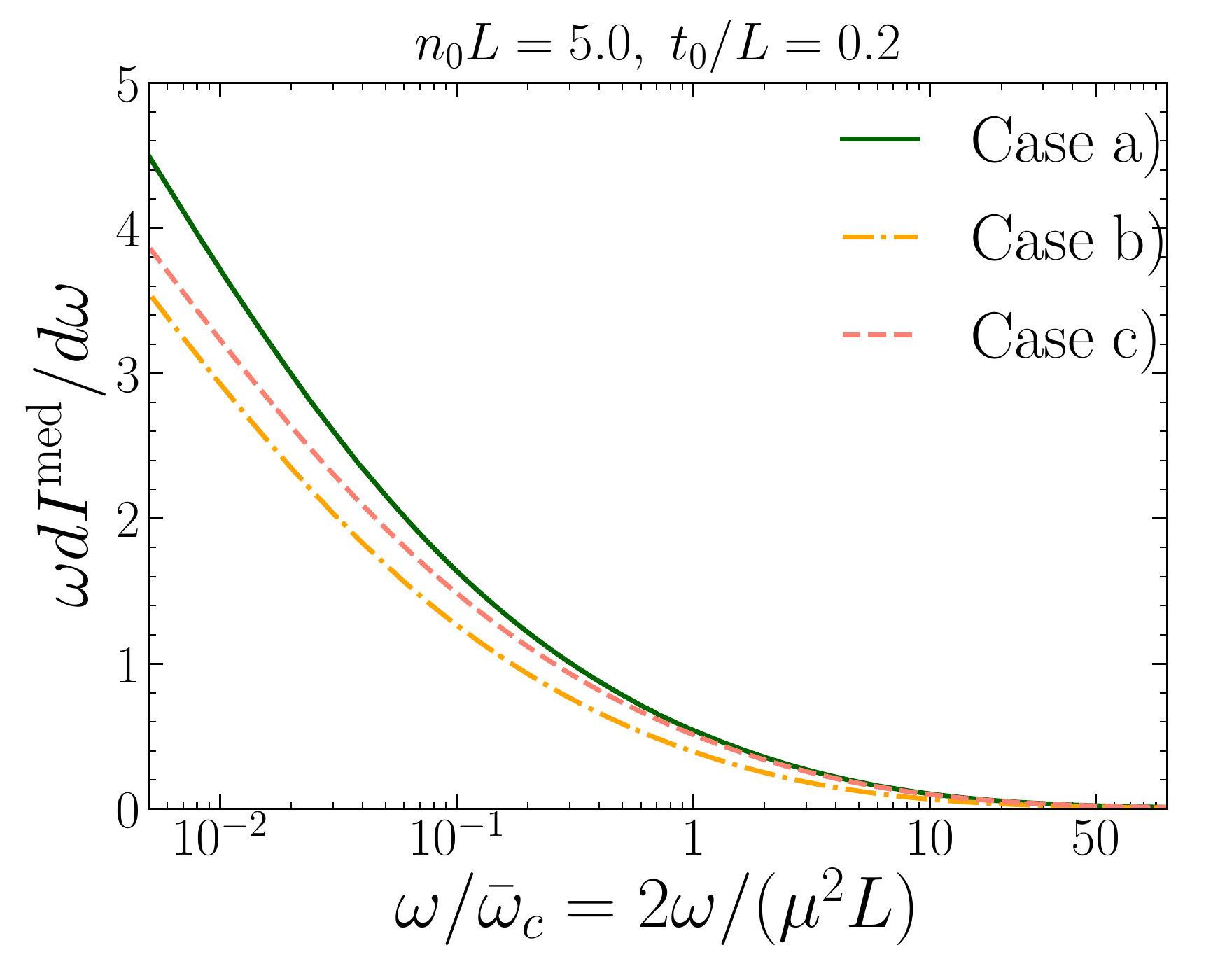}
\caption{Fully resummed medium-induced gluon energy spectrum for the three scenarios: Case a)  (solid green lines), case b) (dashed-dotted orange lines) and case c)  (red dashed lines) as function of $\omega/\bar \omega_c$ for $n_0L=5$ and $t_0/L=0.2$.} 
\label{fig:delay}
\end{figure}

We present in Fig.~\ref{fig:delay} the medium-induced single gluon emission spectrum for the three above-mentioned scenarios, which  can be summarized as follows:
\begin{enumerate}[label=\alph*)] 
\item Parent parton produced at time $0$ propagating through a medium going from $0$ to $L$.
\item Parent parton produced at $t_0$ propagating through a medium going from $t_0$ to $L$.
\item Parent parton produced at $0$ propagating first though vacuum from $0$ to $t_0$ and then through a medium from $t_0$ to $L$.
\end{enumerate}
The scenarios a) and b) are formally equivalent, their only difference being that the former corresponds to a medium of length $L$ while the latter to a medium of length $L-t_0$. Therefore, as one can see in Fig.~\ref{fig:delay}, where we plot the in-medium spectrum for the three scenarios using $t_0/L=0.2$, the result corresponding to the longer medium (case a, green solid) is larger. Scenarios b) and c) correspond to media with the same length, but the latter takes into account extra medium-induced radiation with respect to case b) arising from the interference between the gluon being emitted in vacuum in the amplitude and in medium in the complex amplitude (see Fig.~\ref{fig:cases}). This extra contribution gives rise to a larger spectrum in case c) than in case b), as it can be seen in Fig.~\ref{fig:delay}.

\section{Conclusions}
In this work we make use of the exact calculation of the in-medium single gluon emission spectrum including all multiple scatterings from \cite{Andres:2020vxs} to gain insight into the radiation process. First, we show that coherence effects among multiple scatterings are fundamental to properly describe the radiation process both for low and intermediate gluon energies ($\omega < \bar  \omega_c$). Then, we compute the all-order medium-induced radiation spectrum when a time delay between the hard process that creates the hard parton and the medium production is set. We present preliminary results on the impact of this delay on the spectrum for a static medium. Given the recently proven sensitivity of jet quenching observables to the initial stages  \cite{Andres:2019eus}, performing a thorough analysis of the impact of this delay for dynamic media would be very interesting and will be left for future work.

\acknowledgments{This work was supported  by Ministerio de Ciencia e Innovaci\'on of Spain under project FPA2017-83814-P; Unidad de Excelencia Mar\'{\i}a de Maetzu under project MDM-2016-0692; Xunta de Galicia
under project ED431C 2017/07; Conseller\'{\i}a de Educaci\'on, Universidade e Formaci\'on Profesional as Centro de Investigaci\'on do Sistema universitario de Galicia
(ED431G 2019/05); European Research Council under project ERC-2018-ADG-835105 YoctoLHC; and FEDER.
C.A. was supported through H2020-MSCA-IF-2019 893021 JQ4LHC. L.A. was supported by OE-Portugal, FCT under contract DL57/2016/CP1345/CT0004. M.G.M. was supported by Ministerio de Universidades of Spain through the National Program FPU (grant number FPU18/01966).}

\bibliographystyle{JHEP}
\bibliography{bibliography}

\providecommand{\href}[2]{#2}\begingroup\raggedright\begin{thebibliography}{1}

\bibitem{Andres:2020vxs}
C.~Andres, L.~Apolin\'ario and F.~Dominguez, \emph{{Medium-induced gluon
  radiation with full resummation of multiple scatterings for realistic
  parton-medium interactions}},
  \href{https://doi.org/10.1007/JHEP07(2020)114}{\emph{JHEP} {\bfseries 07}
  (2020) 114} [\href{https://arxiv.org/abs/2002.01517}{{\ttfamily
  2002.01517}}].

\bibitem{Andres:2020kfg}
C.~Andres, F.~Dominguez and M.~Gonzalez~Martinez, \emph{{From soft to hard
  radiation: the role of multiple scatterings in medium-induced gluon
  emissions}}, \href{https://doi.org/10.1007/JHEP03(2021)102}{\emph{JHEP}
  {\bfseries 03} (2021) 102}
  [\href{https://arxiv.org/abs/2011.06522}{{\ttfamily 2011.06522}}].

\bibitem{Mehtar-Tani:2019tvy}
Y.~Mehtar-Tani, \emph{{Gluon bremsstrahlung in finite media beyond multiple
  soft scattering approximation}},
  \href{https://doi.org/10.1007/JHEP07(2019)057}{\emph{JHEP} {\bfseries 07}
  (2019) 057} [\href{https://arxiv.org/abs/1903.00506}{{\ttfamily
  1903.00506}}].

\bibitem{Andres:2019eus}
C.~Andres, N.~Armesto, H.~Niemi, R.~Paatelainen and C.A.~Salgado, \emph{{Jet
  quenching as a probe of the initial stages in heavy-ion collisions}},
  \href{https://doi.org/10.1016/j.physletb.2020.135318}{\emph{Phys. Lett. B}
  {\bfseries 803} (2020) 135318}
  [\href{https://arxiv.org/abs/1902.03231}{{\ttfamily 1902.03231}}].

\end{thebibliography}\endgroup

\end{document}